\documentclass[letterpaper, 10 pt, conference]{ieeeconf}  
\IEEEoverridecommandlockouts                              
\usepackage{cite}
\usepackage{amsmath,amssymb,amsfonts}
\usepackage{algorithmic}
\usepackage{graphicx}
\usepackage{textcomp}
\usepackage{xcolor}
\usepackage{tabularx}
\usepackage[top=20.3mm, bottom=21mm, left=17mm, right=17mm]{geometry}
\def\BibTeX{{\rm B\kern-.05em{\sc i\kern-.025em b}\kern-.08em
    T\kern-.1667em\lower.7ex\hbox{E}\kern-.125emX}}

\title{\LARGE \bf
	Zero Dynamics Attack Detection and Isolation in Cyber-Physical Systems with Event-triggered Communication
}
\author{Ali Eslami and Khashayar Khorasani
	\thanks{Ali Eslami {\tt\small (a\_esla@live.concordia.ca)} and Khashayar Khorasani {(\tt\small kash@ece.concordia.ca)} are with the Department of Electrical and Computer Engineering, Concordia University, Montreal, Quebec, Canada, H3G 1M8.}
}
\begin{document}
	\maketitle
	\thispagestyle{plain}
	\pagestyle{plain}
\begin{abstract}
This paper investigates the problem of Zero Dynamics (ZD) cyber-attack detection and isolation in Cyber-Physical Systems (CPS). By utilizing the notion of auxiliary systems with event-based communications, we will develop a detection mechanism capable of detecting and isolating the ZD cyber-attack even when the attackers have full knowledge of the dynamics of the auxiliary system and can launch False Data Injection (FDI) attacks on all the communication channels. More specifically, we will utilize a self-triggering rule for the communication channels connecting the auxiliary system with the Command \& Control (C\&C) center, leveraging its properties to detect the ZD cyber-attack. Finally, the effectiveness and capabilities of our approach are verified and demonstrated through simulation case studies.
\end{abstract}
\section{Introduction}
\label{sec:Introduction}
In recent years, Cyber-Physical Systems (CPS) have received considerable attention due to their numerous benefits.  Given that CPS has many applications in safety critical systems and infrastructure, securing these systems have become an important challenge for researchers. The security of CPS has been investigated extensively in the literature in the past decade (e.g. \cite{back2017enhancement,baniamerian2021special,hoehn2016detection,celi2022data}). Detecting cyber-attacks is an important aspect in Security of CPS and among the different types of cyber-attacks, Zero Dynamics (ZD) cyber-attack is one of the most powerful and impactful. In this type of cyber-attack, the attacker has extensive knowledge of the system dynamics and possesses the necessary resources to execute a False Data Injection (FDI) attack through the input communication channel. In ZD cyber-attack, the attacker injects false information into the plant’s input channel to excite the system’s zeros. If the system has minimum-phase zeros, the attack's impact diminishes over time, eventually reaching zero. However, if the system has non-minimum-phase zeros, the system states will diverge to infinity while the output remains unchanged.

The authors in \cite{teixeira2012attack} were among the first to describe this type of cyber-attack, investigating the requirements for launching ZD cyber-attacks. Furthermore, Sampling ZD cyber-attack has been investigated where the zeros of the system are due to the sampling period of the sampler \cite{yuz2014sampled}. It has also been shown that attackers do not require exact knowledge of the model to launch ZD cyber-attack, and a robust ZD cyber-attack with uncertain knowledge of the system has been proposed in \cite{shim2016yet}. The ZD cyber-attack in delayed systems has also been investigated in \cite{baniamerian2021special}.

For detection of ZD cyber-attack, \cite{back2017enhancement} addresses the problem of detecting sampling ZD cyber-attack by using a generalized hold instead of a zero-order-hold. This approach is further analyzed in \cite{kim2020neutralizing} and it is mentioned that by assigning sampling zeros to stable locations, the effects of sampling ZD cyber-attack will be neutralized.
A modulation matrix and an observer are also proposed in \cite{hoehn2016detection} to address the problem of detecting both ZD cyber-attack and covert attacks. In this approach, it is assumed that the attacker does not know the parameters of the modulation matrix and the modulation matrix is changed periodically to avoid parameter identification by the attacker.
Another approach that is presented in \cite{baniamerian2020monitoring} is to use an auxiliary system alongside the plant. Then, an auxiliary channel is used to send the auxiliary system's output through the communication channels to the Command and Control (C\&C) side. Observers are then employed in C\&C to detect ZD cyber-attack while assuming that the auxiliary channel is not under cyber-attacks.
 
Event-based communications and control in CPS are gaining increasing attention in the cybersecurity domain. In event-based systems, the exchange of information occurs when an event is triggered by the triggering mechanism. One specific type of event-based communications is the self-triggering mechanism \cite{ma2024resilient} where instead of continuously observing the event condition to determine when an event occurs, we determine the next triggering time of the communication channel at the current event time.
In the literature, event-based and self-triggering communication have been used against various types of cyber-attacks, especially Denial of Service (DoS) attacks.

In this paper, we propose and develop a detection and isolation mechanism for the Zero Dynamics (ZD) cyber-attack in CPS, motivated by critical applications that suffer from existence of non-minimum phase zeros, such as in the flight control of a fighter aircraft \cite{stevens2015aircraft}. 
We first develop an auxiliary system alongside the plant and then construct an observer for the state of the auxiliary system at the C\&C center. In order to reduce the communication bandwidth usage, we consider event-based communication for the output channel of the plant and a self-triggering communication for the output of the auxiliary system. We define three residuals, namely the first captures discrepancies between the auxiliary system's output and its estimate, and the second captures discrepancies between the auxiliary channel's triggering times and the triggering times computed at the C\&C center. Finally, the third one is utilized to help us in isolating ZD cyber-attack. The main advantage of our approach as compared to the existing methods (e.g., \cite{baniamerian2020monitoring}) is that our detection and isolation mechanism can detect and isolate the ZD cyber-attack even when the attacker can launch cyber-attacks on all communication channels and has complete knowledge of both the plant and the auxiliary system dynamics. This consideration allows for our detection and isolation approach to be much more resilient to attackers having access high levels of resources. Therefore, the contributions of the paper can be summarized as follows, namely:
\begin{enumerate}
	\item Developed a novel detection mechanism against ZD cyber-attacks with attackers capable of launching cyber-attacks on all the communication channels.
	\item Designed an isolation mechanism to isolate ZD cyber-attacks from other types of cyber-attacks.
	\item Excluded the possibility of Zeno behavior in our system.
\end{enumerate}

The remainder of this paper is organized as follows: In Section \ref{sec:Problem Formulation}, we provide background information on ZD cyber-attack and discuss the problem formulation. In Section \ref{sec:CyberAttack Detection Mechanism}, we present our detection mechanism and residuals. In Section \ref{sec:Numerical Case Studies}, we provide numerical case studies and in Section \ref{sec:Conclusion}, we conclude the paper.
\section{Background \& Problem Formulation}
\label{sec:Problem Formulation}
\subsection{Zero Dynamics (ZD) Cyber-Attck}
\label{subsec:ZDA}
Consider the following linear time-invariant system:
\begin{align}
	\label{eq:ZDAsystemBackground}
	\begin{split}
		\dot{x}(t)&=Ax(t)+Ba_{u}(t)\\
		y(t)&=Cx(t)
	\end{split}
\end{align}
where $x(t)\in 	\mathbb{R}^n$,  $y(t) \in \mathbb{R}^p$, and $a_{u}(t) \in \mathbb{R}^m$ denote the state, output, and the cyber-attack signal on the input communication channels, respectively. Moreover, consider the Rosenbrock system matrix given by:
\begin{align}
	\label{eq:Rosenbrock Matrix}
	P(s)=\begin{bmatrix} sI - A & -B \\ C & 0 \end{bmatrix}
\end{align}
We now state the following definition:

\textbf{Definition 1}: \cite{teixeira2012attack} Consider the system (\ref{eq:ZDAsystemBackground}) and the pair $\begin{bmatrix} x_0 \\ a_0 \end{bmatrix}$ such that, for $x(0)=x_0$ and $a_{u}(t)=a_0 e^{s_0t}$ it follows that $y(t)\equiv 0$, where $s_0$ is the zero frequency of the system (\ref{eq:ZDAsystemBackground}) if $rank(P(s_0))< normrank(P(s))$.

\subsection{System Dynamics}
\begin{figure}
	\begin{center}
		\includegraphics[width=8.6cm]{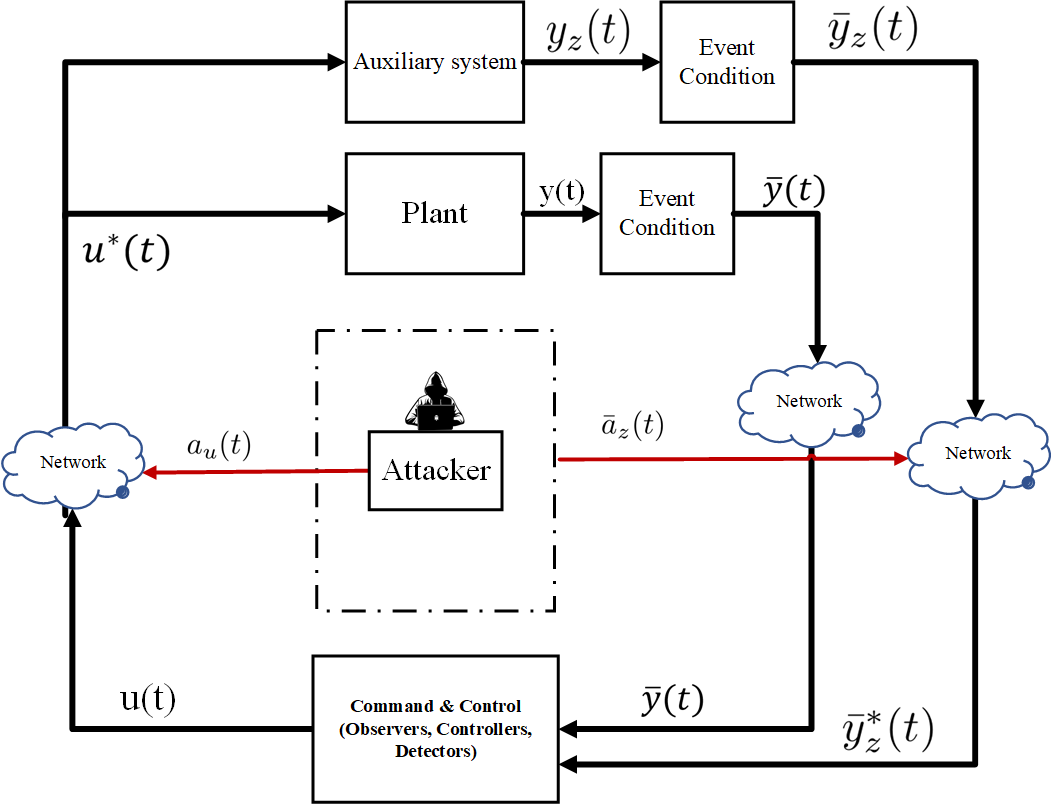}    
		\caption{Overall schematic of the CPS where the input channel is subject to the cyber-attack signal $a_u$ and the auxiliary channel is subject to the cyber-attack signal $\bar{a}_z(t)=a_z(t_j),\forall t \in [t_j,t_{j+1})$. }
		\label{fig:Schematics1}
	\end{center}
\end{figure}

The overall schematics of the CPS are depicted in Fig. \ref{fig:Schematics1} with its plant governed by the following dynamics:
\begin{align}
	\label{eq:System Dynamics}
	\begin{split}
		{{\dot{x}}}(t)&={{A}}{{x}}(t)+{{B}}{u^{*}}(t)\\
		{{y}}(t)&={{C}}{{x}}(t)
	\end{split}
\end{align}
where $x(t) \in 	\mathbb{R}^n$, $y(t) \in \mathbb{R}^p$, and $u^*(t) \in \mathbb{R}^m$ denote the state, measured output, and control input on the plant side, respectively, where $A$, $B$, and $C$ are system matrices of appropriate dimensions. It should be noted that although noise is not included in our mathematical formulation, it will be considered in the simulations conducted in the Section \ref{sec:Numerical Case Studies}.

The control input at the plant side is the latest received input from the C\&C side and can be under cyber-attack, namely:
\begin{equation}
\label{eq:AttackOnU}
u^*(t)=u(t)+a_u(t)
\end{equation}
where $a_u(t)$ denotes the cyber-attack signal on the input channel that was provided in the Definition 1.

In this paper, we have considered the output channels to be event-triggered communication channels. The event condition (also known as the triggering condition or the event triggering condition) is considered as:
\begin{align}
	\label{eq:EventConditionOutput}
	e_{y}^{T}(t)e_{y}(t) < \sigma g(t) 
\end{align} 
where $e_{y}(t)=y(t_k)-y(t)$ denotes the output event error with $t_k$ denoting the latest event time of the output channel and $k \in \mathbb{N}$ with $\mathbb{N}$ denoting the set of natural numbers, $0<\sigma<1$ a positive constant and $g(t)$ a dynamic variable that is governed by the following equation:
\begin{align}
	\label{eq:DynamicVariableG}
	\dot{g}(t)=-c_1 g(t) - c_2  \| e_y(t) \|^2+ \varepsilon
\end{align}
with $c_1>1$, $0<c_2<1 $, and $\varepsilon$ positive constants.

In the event-based communication channels, the event condition (\ref{eq:EventConditionOutput}) is observed continuously and whenever this condition is violated (at event time $t_k$, specifically when $e_{y}^{T}(t_k)e_{y}(t_k)= \sigma g(t_k)$), the output of the system is sent through communication channels and the event error $e_y$ will be updated to zero (since the output at the event time $y(t_k)$ will be updated to $y(t)$). This ensures that the event condition (\ref{eq:EventConditionOutput}) is satisfied again. Consequently, the next event time (i.e. the next time the output will be sent through the communication channels denoted by $t_{k+1}$) will be the next time that the event condition is violated again and this process repeats each time the event condition is violated.

Note that although the triggering condition (\ref{eq:EventConditionOutput}) is not used directly in our detection mechanism, it plays a key role in Theorem 1 where we show stability of the system in cyber-attack free scenario. Furthermore, it is also utilized in Lemma 3 in order to provide an isolation mechanism to isolate the ZD cyber-attack from other types of cyber-attacks. Furthermore, its dynamic threshold capability adapts over time, reducing triggering frequency and bandwidth usage as compared to static methods (refer to \cite{ruan2020observer}, Remark 2).

\textbf{Lemma 1}: \cite{ruan2020observer} The dynamic variable $g(t)$, if initialized as $g(0) > \frac{\varepsilon}{c_1}$, will satisfy the following condition :
\begin{align}
	g(t) > \frac{\varepsilon}{c_1+\sigma c_2 }>0
\end{align}

\textbf{Remark 1}: Lemma 1 ensures that the dynamic variable $g(t)$ remains positive. We utilize this property when developing the Lyapunov function candidate in Theorem 1.

We define the control input signal $u(t)$ as follows:
\begin{align}
	\label{eq:Controller}
	u(t) = K y (t_k)
\end{align}
where $K$ is the controller gain that needs to be designed. 

\textbf{Assumption 1}: $(A,B)$ is controllable and $(A,C)$ is observable.

Our objective is to detect the Zero Dynamics (ZD) cyber-attack in the Command \& Control (C\&C) center. It is important to note that in ZD cyber-attacks, the attacker does not need to launch a cyber-attack on the output communication channel to remain stealthy. This is due to the fact that, in the ZD cyber-attacks, the effects of cyber-attacks on the input channel ($a_u$) are not reflected in the output ($y$), as discussed in the definition of the ZD cyber-attack in the Section \ref{subsec:ZDA}. This is also shown in Fig. \ref{fig:Schematics1} where the attacker does not need to inject any signal to the output channel.

Due to the reasons mentioned above, to develop our ZD cyber-attack detection mechanism, we first consider the following auxiliary system along side the plant:
\begin{align}
	\label{eq:AuxDynamics}
	\begin{split}
	\dot{z}(t) &= A_z z (t) + B_z u^* (t)\\
	y_z(t) &= C_z z(t)
	\end{split}
\end{align}
where $z(t)$ and $y_z(t)$ denote the state and output of the auxiliary system. The auxiliary system (\ref{eq:AuxDynamics}) is designed to have no zero dynamics.

\textbf{Remark 2}: The auxiliary system is not meant to deceive attackers similar to honeypots methods, but rather to aid in detecting ZD cyber-attack by ensuring that the auxiliary system output is influenced by the cyber-attack signal $a_u$.  Note that the output of the plant will not be affected by the ZD cyber-attack since this type of cyber-attack excites the zeros of the system. Therefore, while the output of the system is unchanged when the system is subject to ZD cyber-attack, the output of the auxiliary system will be affected by ZD cyber-attack since the auxiliary system (\ref{eq:AuxDynamics}) is designed to not have any zero dynamics. Considering auxiliary system is a common methodology in the literature (mostly utilized against covert attacks, e.g., \cite{griffioen2020moving,schellenberger2017detection}) and practical due to the concurrent processing capabilities of modern CPS systems.

The output of the auxiliary system ($y_z$) is sent to the C\&C center through a specific type of event-based communication, namely self-triggering communication. In self triggering communication, the next event time of the communication channel is determined at the current event time. Although the self-triggering communication is a specific type of event-based communication, it is generally more conservative, resulting in more frequent event triggerings. However, the advantage of self-triggering communication is that it eliminates the need to continuously monitor the event condition to determine the event triggerings. In this paper, we have chosen self-triggering communication to utilize its capabilities for the ZD cyber-attack detection, which will be discussed in the next section.

We will first provide the event condition from which we will obtain the self-triggering communication rule. The corresponding event condition for the auxiliary channel is considered as follows,
\begin{align}
	\label{eq:EventAux}
	\| e_z (t) \| < \frac{1}{\sqrt{4+4\delta^2}}(\| y_z(t_j) \| + \sqrt{2\varepsilon_2})
\end{align}
where $e_{z}(t)=y_{z}(t_j)-y_z(t)$ denotes the event error of the auxiliary system output with $t_j$ the latest event time of the auxiliary system where $j \in \mathbb{N}$ with $\mathbb{N}$ the set of natural numbers and $\delta>1$ and $\varepsilon_2$ positive constants. Furthermore, note that $y_{z}(t_j)$ represents the output that was sent to the C\&C center at the latest event time  of the auxiliary system ($t_{j}$). As stated earlier, we will use the event condition (\ref{eq:EventAux}) to derive the self-triggering rule that will be provided subsequently in this section. 

Given that the output of the auxiliary system is sent through the communication network, it can be subject to cyber-attacks:
\begin{align}
	\label{eq:AttackonZ}
	y_{z}^{*}(t_{j})=y_{z}(t_{j})+a_{z}(t_{j})
\end{align}
where $y_{z}^{*}(t_j)$ denotes the output of the auxiliary system corrupted by the cyber-attack signal $a_{z}(t_j)$, injected at the event times of the auxiliary channel $t_{j}$. It should be pointed out that the cyber-attack signal $a_{z}$ is a False Data Injection (FDI) cyber-attack and its design is dependent on the attackers knowledge about the auxiliary system. We will provide further details on the design procedure of the cyber-attack signal $a_z$ in the Section \ref{sec:CyberAttack Detection Mechanism}.

\textbf{Assumption 2}: The cyber-attack signals $a_u(t)$ and $a_{z}(t)$ are considered to be occurring on the physical layer, i.e. the attacker transmit a signal to be combined with the original signal of the system (e.g. Symbol Flipping \cite{popper2011investigation}).

Let us consider the following observer which is placed at the C\&C center for the auxiliary system, that is
\begin{align}
	\label{eq:ObserverAux}
	\dot{\hat{z}}(t)=A_z \hat{z}(t)+B_z u(t) - L (\bar{y}_{z}^{*}(t) - C_z \hat{z}(t))
\end{align}
where $\hat{z}(t)$ denotes the estimate of the auxiliary system state, $\bar{y}_{z}^{*}(t) = y_{z}^{*}(t_j) = y_z(t_j)+a_{z}(t_j), \forall t \in [t_j,t_{j+1})$ and $L$ denotes the observer gain to be designed.

Let us construct the following augmented state $\eta = \begin{bmatrix} x^T, z^T, \tilde{z}^{T} \end{bmatrix}^T$ with  $\tilde{z}(t) = z(t) - \hat{z}(t)$ which is the estimation error for the observer (\ref{eq:ObserverAux}). Therefore, we can obtain:
\begin{align}
	\label{eq:AugmentedDynamics}
	\dot{\eta}(t) = A_{\eta} \eta(t) + B_\eta e(t) + B_a a(t)
\end{align}
where $e(t) = \begin{bmatrix} e_{y}^{T}(t), e_{z}^{T}(t) \end{bmatrix}^T$, $A_{\eta} =\begin{bmatrix} A+BKC & 0 & 0 \\ B_zKC & A_z & 0 \\  0 & 0 & A_z+LC_z \end{bmatrix} $, $B_\eta = \begin{bmatrix} BK & 0 \\ B_zK & 0 \\ 0 & L \end{bmatrix}$ , $a(t) = \begin{bmatrix} a_{u}^{T}(t), \bar{a}_{z}^{T}(t) \end{bmatrix}^T$, $\bar{a}_{z}(t) = a_{z}(t_j), \forall t \in [t_j,t_{j+1})$, and $B_a = \begin{bmatrix} B & 0\\ B_z & 0 \\ 0 & L \end{bmatrix}$.

To demonstrate that in cyber-attack free scenarios, the augmented dynamics $\eta(t)$ will be uniformly ultimately bounded, we use the following lemma.

\textbf{Lemma 2}: The event condition (\ref{eq:EventAux}) is a sufficient condition for the following condition:
\begin{align}
	\label{eq:EventAuxOriginal}
	e_{z}^{T}(t)e_{z}(t) < \delta y_{z}^{T}(t)y_{z}(t) + \varepsilon_2
\end{align}

\textbf{Proof}: Take the condition (\ref{eq:EventAux}), we can obtain:
\begin{align}
	\begin{split}
		\| e_z(t) \|^2 &< \frac{1}{4+4\delta^2}(\| y_z(t)+e_z(t) \|^2 + 2\varepsilon_2 \\&+ 2\|y_{z}(t)+e_{z}(t)\| \sqrt{2\varepsilon_2})
		\end{split}
\end{align}

Then, by using the well-known inequality $a^2 + b^2 \ge 2ab$, we have
$\| e_z(t) \|^2 < \frac{1}{4+4\delta^2}(2\| y_z(t)+e_z(t) \|^2 + 4\varepsilon_2) \le \frac{1}{4+4\delta^2} (4(\| y_z(t) \|^2 + \| e_z \|^2 )+4\varepsilon_2)  \le \frac{1}{1+\delta^2}(\|y_z(t) \|^2 + \|e_z(t) \|^2+\varepsilon_2)$.
Thus, we obtain:
$\delta ^2 \| e_z(t) \|^2 < \| y_z(t) \|^2 + \varepsilon_2$ or $
	\| e_z(t) \|^2 < \frac{1}{\delta^2} \|y_z(t) \|^2 + \frac{1}{\delta^2} \varepsilon_2$.
Therefore, the event condition (\ref{eq:EventAux}) is a sufficient condition for (\ref{eq:EventAuxOriginal}) since $\delta >1$. \hfill $\blacksquare$

\textbf{Theorem 1}: Under the Assumption 1, the event conditions (\ref{eq:EventConditionOutput}) and (\ref{eq:EventAux}), and given positive definite matrix $P$, the augmented dynamics $\eta (t)$ will be Uniformly Ultimately Bounded (UUB) if the following Linear Matrix Inequality (LMI) is satisfied:
\begin{align}
	\label{eq:LMITheorem}
	\Pi = \begin{bmatrix}
		 \Pi_1 & PB_\eta \\
		* & -(1-c_2)F^T F - F_{2}^{T}F_2
	\end{bmatrix}<0
\end{align}
where $\Pi_1 = -\frac{1}{2}P + \delta F_{3}^{T}C^T C F_3$, $F = \begin{bmatrix}
	I & 0
\end{bmatrix}$, $F_2 = \begin{bmatrix}
0 & I
\end{bmatrix}$, and $F_3 = \begin{bmatrix}
0 & I & 0
\end{bmatrix}$.

\textbf{Proof}. To provide proof, we first note that since the system dynamics (\ref{eq:System Dynamics}) is controllable, and since the design of the auxiliary system ($A_z,B_z,C_z$) and the observer gain $L$ is in our discretion, one is capable of designing $K$ and $L$ to make the matrix $A_\eta$ Hurwitz. This would then imply that there exists a positive definite matrix $P$ such that:
\begin{align}
	\label{eq:LyapunovEarly}
	A_{\eta}^{T}P + PA_{\eta} < -P
\end{align}
We now define our Lyapunov function candidate as $V(t) = \eta^T(t) P \eta(t) + g(t)$. Note that from Lemma 1, it has been shown that $g(t)$ remains positive. Then, by taking the derivative of $V$, we obtain:
\begin{align}
	\dot{V}(t) &= \eta^T(t) (A_{\eta}^T P + PA_{\eta})\eta(t) + 2 \eta^T(t) PB_{\eta}e(t) \\&-c_1 g(t) - c_2 e_{y}^{T}(t)e_{y}(t)+\varepsilon
\end{align}
By adding and subtracting $e_{y}^{T}(t)e_{y}(t)$ and $e_{z}^{T}(t)e_{z}(t)$, we obtain:
\begin{align}
	\label{eq:LyapunovMid1}
	\begin{split}
	\dot{V}(t) &\le \eta^T(t) (A_{\eta}^T P + PA_{\eta})\eta(t) + 2 \eta^T(t) PB_{\eta}e(t) \\&
	-e_{y}^{T}(t)e_{y}(t) + \sigma g(t) - c_{1}g(t) - c_2e_{y}^{T}(t)e_{y}(t)\\&+\varepsilon -e_{z}^{T}(t)e_{z}(t)+\delta y_{z}^{T}(t)y_{z}(t)+\varepsilon_2
		\end{split}
\end{align}
Note that in (\ref{eq:LyapunovMid1}), we have utilized the inequality $e_{y}^{T}(t)e_{y}(t) \le \sigma g(t)$ which is true due to the event condition (\ref{eq:EventConditionOutput}) and the inequality $e_{z}^{T}(t)e_{z}(t) < \delta y_{z}^{T}(t)y_{z}(t)+\varepsilon_2$ which is true due to the event condition (\ref{eq:EventAux}) and the condition (\ref{eq:EventAuxOriginal}) in Lemma 2. 

Furthermore, note that $e_{y}^{T}(t)e_{y}(t) = e^T(t)F^TF e(t)$ with $F = \begin{bmatrix}
	I & 0
\end{bmatrix}$, $e_{z}^{T}(t)e_{z}(t) = e^T(t)F_{2}^{T}F_{2} e(t)$ with $F_2 = \begin{bmatrix}
0 & I
\end{bmatrix}$, and $y_{z}^{T}(t)y_{z}(t) = \eta^{T}(t)F_{3}^{T}C^TCF_{3}\eta(t)$ with $F_{3} = \begin{bmatrix}
0 & I & 0
\end{bmatrix}$. Therefore, we obtain the following:
\begin{align}
	\begin{split}
		\dot{V}(t) &\le \eta^T(t) (A_{\eta}^T P + PA_{\eta})\eta(t) + 2 \eta^T(t) PB_{\eta}e(t) \\&
 + \sigma g(t) - c_{1}g(t) - c_2e^T(t)F^TF e(t)+\varepsilon+\varepsilon_2\\& 	-e^T(t)F^TF e(t)-e^T(t)F_{2}^{T}F_{2} e(t)\\&+\delta \eta^{T}(t)F_{3}^{T}C^TCF_{3}\eta(t)
	\end{split}
\end{align}
In other words, by using equation (\ref{eq:LyapunovEarly}), one can obtain:
\begin{align}
	\begin{split}
		\dot{V}(t) &\le -\frac{1}{2}\eta^T(t)P \eta(t) + (\sigma - c_1)g(t)+\varepsilon + \varepsilon_2 \\&+ 
		\begin{bmatrix}
			\eta \\ e
		\end{bmatrix}^T \Pi \begin{bmatrix}
		\eta \\ e
		\end{bmatrix}
	\end{split}
\end{align}
where $\Pi$ is defined in the LMI (\ref{eq:LMITheorem}). If the LMI (\ref{eq:LMITheorem}) is satisfied, we can then obtain
	$\dot{V}(t) \le -\beta V(t) + \varepsilon_3
$, where $\beta = min\begin{Bmatrix}
	{\frac{1}{2}\lambda_{min}\{P\}, (c_1 - \sigma)}
\end{Bmatrix}$ and $\varepsilon_3 = \varepsilon + \varepsilon_2$.
Therefore, we obtain
$V(t) \le (V(0) - \frac{\varepsilon_3}{\beta})e^{-\beta t} + \frac{\varepsilon_3}{\beta}$.
Since $e^{-\beta t}$ approaches $0$ as $t \to \infty$, there exists a positive number $T$ such that for all $t\ge T$, $V(t) \le \frac{2 \varepsilon_3}{\beta}$. This shows that the Lyapunov function, consisting of the system state $x(t)$, the auxiliary system state $z(t)$, and the estimation error $\tilde{z}(t)$ is uniformly ultimately bounded. This completes the proof of the theorem. \hfill $\blacksquare$

\textbf{Remark 3}: The ultimate bound of the augmented system state $\eta$ is influenced by the event condition thresholds $\varepsilon_1$ and $\varepsilon_2$. Increasing these values reduces the number of event triggerings but enlarges the ultimate bound of $\eta$. Thus, designing the event conditions (\ref{eq:EventConditionOutput}) and (\ref{eq:EventAux}) requires balancing communication bandwidth and system performance. Although the auxiliary system matrices $(A_z, B_z, C_z)$ do not affect the ultimate bound, they do influence the convergence rate of the Lyapunov function. Specifically, these matrices affect $A_\eta$ in the inequality (\ref{eq:LyapunovEarly}), which in turn influences the matrix $P$ that is required to satisfy the LMI (\ref{eq:LMITheorem}). Consequently, the convergence rate $\beta$ of the Lyapunov function depends on $P$.

Now, we turn our focus to the Zeno behavior and we demonstrate that the Zeno behavior will not occur under the event conditions (\ref{eq:EventConditionOutput}) and (\ref{eq:EventAux}). Zeno behavior, a phenomena where infinte number of triggering occur in a finite time, is impractical and incompatible with real systems, as communication devices cannot support such infinite transmissions. The presence of Zeno behavior would invalidate the system's model due to physical and technological constraints. To address this critical challenge, we next show that a positive lower bound exists between the event times, ensuring that event-triggered communications occur at a finite rate.

\textbf{Theorem 2}: Given the event conditions (\ref{eq:EventConditionOutput}) and (\ref{eq:EventAux}) and the results obtained from Theorem 1, there exist positive constants $T_1$ and $T_2$ such that $t_{k+1}-t_{k}>T_1$ and $t_{j+1}-t_{j}>T_2$, i.e. the Zeno behavior will not occur in our system.

\textbf{Proof}: We first start with the triggering condition (\ref{eq:EventAux}). It can be seen that we have:
\begin{align}
	\dot{e_z}=-C_zA_z z(t) - C_z B_z u^*(t)
\end{align}
Therefore, we can obtain:
\begin{align}
	\begin{split}
		\frac{d}{dt} \|e_z(t) \| &= \frac{\|e_{z}^{T}(t)\|}{\|e_z(t)\|} \|\dot{e}_z (t)\| \\ &= \| -C_zA_z z(t) - C_z B_z u^*(t)\|
	\end{split}
\end{align}
By adding and subtracting $C_zA_zz(t_j)$ and noting that $z(t) = z(t_j)+e_z(t)$, we obtain:
\begin{align}
	\begin{split}
		&\frac{d}{dt}\|e_z(t) \| \le \| A_z e_z(t) - C_z A_z z(t_j) - C_z B_z u^*(t) \| \\ &\le \| A_z \| \| e_{z}(t) \| + \| A_z \| \| y_z(t_j) \| + \| C_z B_z \| \| u^*(t) \|
	\end{split}
\end{align}
Since both $z(t_j)$ and $u(t)$ are bounded due to the results of Theorem 1, we have:
\begin{align}
	\label{eq:ZenoLastAux}
	\frac{d}{dt}\| e_z(t) \| \le \| A_z \| \|e_{z}(t) \| + M
\end{align}
where $M =\| A_z \| \| y_z(t_j) \| + \| C_z B_z \| \| u^*(t) \| $. 
Note that $M$ is upper bounded in $[t_{j},t_{j+1})$.
Thus, we obtain:
	$\| e_z (t) \| \le \frac{M}{\|A_z\|} (e^{\|A_z\|(t-t_j)}-1)$.
At the next event time $t_{j+1}$, we have $\| e_z (t_{j+1}) \| = \frac{1}{\sqrt{4+4\delta^2}}(\| y_z(t_{j+1}) \| + \sqrt{2\varepsilon_2})$. Therefore, one can obtain
$
	\| e_z(t_{j+1}) \|= \frac{1}{\sqrt{4+4\delta^2}}(\| y_z(t_{j+1}) \| + \sqrt{2\varepsilon_2}) \ge   \frac{1}{\sqrt{4+4\delta^2}}\sqrt{2\varepsilon_2}
$.
Now, let us denote $M_2 = \frac{1}{\sqrt{4+4\delta^2}}\sqrt{2\varepsilon_2}$, and then by using the above equations, we can obtain
$ M_2 \le \frac{M}{\|A_z\|} (e^{\|A_z\|(t_{j+1}-t_j)}-1)$.
Consequently, one can obtain:
\begin{align}
	t_{j+1}-t_{j} \ge \frac{1}{\| A_z\|} ln(\frac{\|A_z\|M_2}{M}+1)>0
\end{align}
Therefore, $t_{j+1}-t_{j}$ is lower bounded by a positive constant.

The proof for the event condition (\ref{eq:EventConditionOutput}) is similar to the event condition (\ref{eq:EventAux}) and is omitted for the sake of brevity. This completes the proof of the theorem. \hfill $\blacksquare$ 

We now derive the self-triggering rule that is obtained from the event condition (\ref{eq:EventAux}).
Let us denote $s_j=\frac{1}{\sqrt{4+4\delta^2}}(\| y_z(t_j) \| + \sqrt{2\varepsilon_2})$ and consider (\ref{eq:ZenoLastAux}), so that by using the event condition (\ref{eq:EventAux}), we can obtain:
\begin{align}
	\begin{split}
	\frac{d}{dt}\|e_{z}(t) \| \le \| A_z \| \|e_z(t) \| + M \le \| A_z\| s_j + M
	\end{split}
\end{align}
Let us denote $\omega_j = \| A_z\| s_j + M$. Then, one can obtain:
\begin{align}
	\| e_z(t) \| \le \int_{t_j}^{t_{j+1}}\frac{d}{dt}\|e_z(t)\|dt \le \omega_j (t_{j+1}-t_{j})
\end{align}
The next event time $t_{j+1}$ is the least time it takes for the event error $e_z(t)$ to go from $0$ to $s_j$.
Therefore, we choose the next event time of the auxiliary channel to be:
\begin{align}
	\label{eq:Self triggering}
	t_{j+1}=t_j + \frac{s_j}{\omega_j}
\end{align}
Note that the self triggering rule (\ref{eq:Self triggering}) satisfies the event condition (\ref{eq:EventAux}) and therefore, the results obtained in Theorem 1 and Theorem 2 will be valid with the self triggering rule (\ref{eq:Self triggering}).

Furthermore, it should be noted that the next triggering time is computed when an event is triggered on the auxiliary channel or when an event is triggered on the output channel. Whenever an event is triggered on the output channel, this will results in the control input signal $u$ being updated, which will change $\omega_{j}$. Therefore, since $\omega_j$ has changed, the triggering time needs to be computed again. 
\section{Cyber-Attack Detection Scheme} 
\label{sec:CyberAttack Detection Mechanism}
In the self-triggering mechanism (\ref{eq:Self triggering}), the next triggering time is computed on the plant side using $s_j$ and $\omega_j$. Let us consider the same auxiliary system in the C\&C center:
\begin{align}
	\begin{split}
	\dot{z}_C(t)&=A_z z_C(t)+B_z u(t) \\
	y_{z}^{C}(t)&=C_zz_C(t)
	\end{split}	
\end{align}

 We now compute the next triggering time of the auxiliary channel using the information available at C\&C center,
$t_{i+1}^{C} = t_{i}^{C}+\frac{s_{i}^{C}}{\omega_{i}^{C}}$,
where $s_{i}^{C}=\frac{1}{\sqrt{4+4\delta^2}}(\| y_{z}^{C}(t_i) \| + \sqrt{2\varepsilon_2})$, and $\omega_{i}^{C}= \|A_z\| s_{i}^{C}+M^{C}$ with $M^{C} = \| A_z \| \| y_{z}^{C}(t_i) \| + \| C_z B_z \| \| u(t) \| $. The information used to compute the next triggering time of the auxiliary channel in the C\&C center ($t_{i+1}^{C}$) is the same as the information utilized to compute the next triggering time on the plant side ($t_{j+1}$) in \textbf{cyber-attack free} scenarios, if the initial conditions of $z$ and $z_C$ satisfy $z(0)=z_C(0)$. Furthermore, similar to the self-triggering rule in (\ref{eq:Self triggering}), $t_{i+1}^{C}$ is computed at the current event time $t_{i}^{C}$ or when an event is triggered on the output channel (causing $\omega_{i}^{C}$ to be updated). 

We are now ready to construct our residuals. 
We will construct two residuals in the C\&C center in order to detect the ZD cyber-attack. Let us first construct the following residual that captures the discrepancy between the output of the auxiliary system and its estimation:
\begin{align}
	\label{eq:Res_z}
	res_z (t) = \bar{y}_{z}^{*}(t) - C_z \hat{z}(t)
\end{align}
Then, a cyber-attack alarm is raised if the residual $res_z(t)$ exceeds a predefined threshold $\gamma_z$. In general, the threshold is chosen based on the noise levels and the acceptable false alarm rates.
However, in our case, the threshold $\gamma_z$ is chosen not only based on the noise levels, but also on the ultimate bound of the estimation error that is provided in the Theorem 1. It should be noted that the considered threshold will not prevent us from detecting the ZD cyber-attacks since ZD cyber-attacks will cause the states of the system to go to infinity. However, this will increase the time it takes for the residual to exceed the threshold. Therefore, when designing the event conditions (\ref{eq:EventConditionOutput}) and (\ref{eq:EventAux}), a trade-off needs to be considered on the utilized communication bandwidth and the detection delay of the residual in (\ref{eq:Res_z}).

Let us now construct the second residual that tries to capture the discrepancy between the triggering time on the plant side with the computed triggering time in the C\&C center, namely
\begin{align}
	\label{eq:Res_t}
	dis_t = t_{i}^{C} - t_{j}
\end{align}
It should be pointed out that in the C\&C center, $t_{j+1}$ is obtained by using the actual arrival time of the information, i.e. instead of computing $t_{j+1}$, our residual constructs $t_{j+1}$ by using a clock and observing when $y_z$ is arrived, the time would be assigned to $t_{j+1}$. Therefore, the arrival time of the information ($t_{j+1}$) is compared with the computed time ($t_{i+1}^{C}$) and if the discrepancy between them exceeds a predefined threshold (for this residual, the threshold is considered to be $0$), a cyber-attack alarm is raised.

We now have the following theorem that deals with ensuring that our residuals (\ref{eq:Res_z}) and (\ref{eq:Res_t}) will be capable of detecting the ZD cyber-attack.

\textbf{Theorem 3}: Given the system dynamics (\ref{eq:System Dynamics}), the ZD cyber-attack provided in the Definition 1 will be detected by the residuals (\ref{eq:Res_z}) and (\ref{eq:Res_t}).

\textbf{Proof}: For the proof of this theorem, we show that the residuals exceed their thresholds when the system is subject to the ZD cyber-attack. We consider two scenarios, namely: a) the attacker lacks the knowledge of the auxiliary system dynamics, and b) the attacker has full knowledge of the auxiliary system dynamics and can design $\bar{a}_z$ to remove the effects of the cyber-attack signal $a_u$ from the auxiliary channel.

a) In this scenario, the attacker cannot design $\bar{a}_z$ to remove the effects of the cyber-attack from the auxiliary channel and this will result in the residual $res_z$ to go beyond the threshold, which leads to detection by the residual $res_z$ in (\ref{eq:Res_z}).

b) In this scenario, the attacker can design $a_z$ as follows:
\begin{align}
	\label{eq:CovertAz}
	\begin{split}	
	\dot{x}_{a}(t)&=A_z x_{a}(t)+Ba_u(t) \\&
	a_z(t) = -C_zx_{a}(t)
\end{split}
\end{align} 
Then, at the event times $t_j$, $a_z$ is injected to the auxiliary channel. This is similar to the covert attack design and interested readers can refer to our previous work in \cite{eslami2024event} for details on how such cyber-attacks remain stealthy from the observer-based residuals similar to (\ref{eq:Res_z}). Note that $a_z$ is a special case of covert attacks where the cyber-attack signal $a_z$ is injected to the output of the auxiliary system $y_z$, instead of  being injected to the output of the plant $y$ similar to general covert attacks. Then, in our formulation, while the attacker will remain stealthy from $res_z$, we also have $dis_t$ in (\ref{eq:Res_t}). In order for the attacker to remain stealthy from the residual in (\ref{eq:Res_t}), the attacker needs to ensure that $\frac{s_{j}}{\omega_j}=\frac{s_{i}^{C}}{\omega_{i}^{C}}$. This implies that the attackers need to ensure the following condition:
\begin{align}
	\label{eq:AttackStealthyRequirement}
	\begin{split}
	\frac{\frac{1}{\sqrt{4+4\delta^2}}(\| y_z(t_j) \| + \sqrt{2\varepsilon_2})}{\| A_z\| s_j + \| A_z \| \| y_z(t_j) \| + \| C_z B_z \| \| u(t)+a_u(t) \|} = \\ \frac{\frac{1}{\sqrt{4+4\delta^2}}(\| y_{z}^{C}(t_i) \| + \sqrt{2\varepsilon_2})}{\|A_z\| s_{i}^{C}+\| A_z \| \| y_{z}^{C}(t_i) \| + \| C_z B_z \| \| u(t) \| }
\end{split}
\end{align}
However, $a_u$ is designed to act as a ZD cyber-attack with its formula given in the Definition 1 ($a_u(t)=a_0 e^{s_0t}$). Therefore, the attacker cannot design $a_u$ to satisfy equation (\ref{eq:AttackStealthyRequirement}). Hence, the ZD cyber-attack cyber-attack will cause the residual in (\ref{eq:Res_t}) to exceed the threshold and this completes the proof of the theorem. \hfill $\blacksquare$

\textbf{Remark 4}: Our proposed detection architecture is specially applicable to cyber-attacks that occur in the physical layer. For example, one of the FDI attacks on the physical layer is Symbol Flipping \cite{popper2011investigation}, where the attacker transmits a signal to be combined with the original signal of the system, modifying the values of the received signal. In this framework, even if the attacker is aware of the event-triggering policy of the system, our detection mechanism will be capable of detecting the ZDA under any FDI cyber-attack on the auxiliary channel. Furthermore, while the detection of DoS cyber-attacks has been extensively studied in the literature (e.g., \cite{gniewkowski2020overview}), the detection of ZD cyber-attacks in scenarios where communication channels are simultaneously vulnerable to both false data injection (FDI) and DoS cyber-attacks remains as our future works.

\textbf{Remark 5}: While we have considered a threshold of zero for the discrepancy (\ref{eq:Res_t}), in practical scenarios, the communication protocols have latency and the threshold should be equal to the latency of the specific communication protocol. For example, CAN Bus protocol has a latency of $(50-200) \mu s$ leading us to consider $200\mu s$ as the threshold for (\ref{eq:Res_t}). Furthermore, in the environments where there exists time-varying bounded delay, our detection mechanism can be extended to those environments by utilizing the statistical properties of the delay such as average and standard deviation. Then, by utilizing approaches similar to $\chi^2$, since the ZD cyber-attack will change the statistical properties of the discrepancy in (\ref{eq:Res_t}), this information can be utilized to extend our proposed detection mechanism to delayed environments. 

before presenting our isolation mechanism, we have the following algorithm to design the auxiliary system matrices $(A_z,B_z,C_z)$ and the corresponding gains $(K,L)$:
\begin{enumerate}
	\item Choose $A_z=Diag(\lambda_0)_{m\times m}$, $B_{z}=I_m$, $C_{z}=I_m$ where $Real\{\lambda_0\}<0$.
	\item Choose $K$ and $L$ to make $A_{\eta}$ Hurwitz.
\end{enumerate}

In order to isolate the ZD cyber-attack from other types of cyber-attacks, let us first construct the following observer at the C\&C center, namely
$\dot{\hat{x}}(t)=A\hat{x}(t)+Bu(t)-L_{2}(y(t_k)-C\hat{x}(t))$,
where $\hat{x}$ denotes the estimation of the system state $x$ and $L_2$ denotes the observer gain. Then, one can obtain the following estimation error dynamics for the system state $x(t)$ and the observer $\hat{x}(t)$, that is
$\dot{\tilde{x}}(t)=(A+L_2C)\tilde{x}(t)+L_2e_y(t)
$.
We now have the following lemma that ensures the estimation error will be UUB.

\textbf{Lemma 3}: The estimation error $\tilde{x}(t)$ will be Uniformly Ultimately Bounded (UUB) under the cyber-attack free scenario if there exists a positive definite matrix $P$ such that the following Linear Matrix Inequality is satisfied, namely:
\begin{align}
	\label{eq:Lemma3}
\begin{bmatrix}
-\frac{1}{2}P & PL_2 \\
* & -(1+c_2)I
\end{bmatrix}<0.
\end{align}

\textbf{Proof}: Similar to the proof in Theorem 1, consider the Lyapunov function candidate as $V(t)=\tilde{x}(t)P\tilde{x}(t)+g(t)$. Then, by denoting $\tilde{A}=A+L_2C$, one can obtain:
\begin{align}
	\dot{V}=\tilde{x}^T(\tilde{A}^TP+P\tilde{A})\tilde{x}+2\tilde{x}PL_2e_y+\dot{g}
\end{align}
Then, given Assumption 1 and by adding and subtracting $e_{y}^{T}e_y$, one can obtain:
\begin{align*}
	\dot{V}&\le -\tilde{x}^TP\tilde{x}+2\tilde{x}^TPL_2e_y-e_{y}^{T}e_y + \sigma g -c_1g-c_2e_{y}^{T}e_y + \varepsilon\\&
     \le \begin{bmatrix}
     	\tilde{x} \\ e_y
     \end{bmatrix}^T \begin{bmatrix}
     	-\frac{1}{2}P & PL_2 \\
     	* & -(1+c_2)I
     \end{bmatrix} \begin{bmatrix}
     	\tilde{x} \\ e_y
     \end{bmatrix} -\frac{1}{2}\tilde{x}^TP\tilde{x}\\&-(c_1-\sigma)g+\varepsilon		
\end{align*}
Then, by satisfying LMI (\ref{eq:Lemma3}), similar to the results in Theorem 1 we can obtain:
\begin{align}
	\dot{V}\le -\beta_2V(t)+\varepsilon
\end{align}
where $\beta=min\{\frac{1}{2}P,(c_1-\sigma)\}$. The rest of the proof is similar to Theorem 1. This completes the proof of the Lemma.
\hfill $\blacksquare$

In order to isolate ZD cyber-attacks from non-ZD cyber-attacks, we first form the following residual that is denoted as $res_x(t)$, namely
$res_x(t)=y(t_k)-C\hat{x}(t)$.
Whenever the residual goes beyond a predefined threshold $\gamma_x$ (which is chosen similar to the threshold $\gamma_2$ that is chosen for $res_z$), the cyber-attack alarm is raised.
The following logic can be utilized for isolation of ZD cyber-attacks:
\begin{itemize}
	\item if $res_x\le\gamma_x$, $res_z\le \gamma_z$, $dis_t=0$ $\to \nexists$ a cyber-attack.
	\item if $res_x>\gamma_x$, $res_z>\gamma_z$, $dis_t>0$ $\to \exists$ a non-ZD cyber-attack.
	\item if $res_x\le \gamma_x$, $res_z>\gamma_z$, $dis_t>0$ $\to \exists$ a ZD cyber-attack.
	\item if $res_x\le\gamma_x$, $res_z\le\gamma_z$, $dis_t>0$ $\to \exists$ a ZD cyber-attack and the auxiliary channel is subject to cyber-attacks and the attacker has obtained full knowledge of the auxiliary system dynamics.
\end{itemize}
\section{Simulation Case Studies}
\label{sec:Numerical Case Studies}
In this section, we consider two different case studies. In the first case study, a linearized model of the quadruple tank process \cite{baniamerian2020monitoring} that is subject to a ZD cyber-attack is considered. In the second case study, we consider the throttle to speed transfer function of an aircraft (\cite{stevens2015aircraft}, Example 3.8-5) subject to ZD cyber-attacks.

\textbf{Case I}: The quadruple tank process has a non minimum phase zero at $s_0=0.01273$ and for $x_0=[0, 0, -0.63597, 0.618476]^T$, we have $a_0=[0.33778, -0.314538]^T$. In order for our results to be closer to practical scenarios, we have also considered a white noise with power of $0.01$. The threshold $\gamma_z$ is considered as $0.01$ and the event condition constants are $\delta=20$, $\varepsilon=\varepsilon_2=0.0001$, $\sigma=0.1$, $c_1=10$, $c_2=0.5$. As stated above, by increasing these constants, the utilized communication bandwidth will be reduced while the ultimate bound of the Lyapunov function will be increased.
By utilizing the algorithm provided to design the auxiliary system, The matrices $(A_z,B_z,C_z)$ are chosen as $A_z=-I_2$, \( B = I_2 \), and $C_z=I_2$. The gains $K$ and $L$ are then chosen as $K=\begin{bmatrix}
	    0.0094 &   0.0295\\
	-0.0042 &   0.0344
\end{bmatrix}$ and $L=-9\times I_2$. 

\begin{figure}
	\begin{center}
		\includegraphics[width=8.6cm]{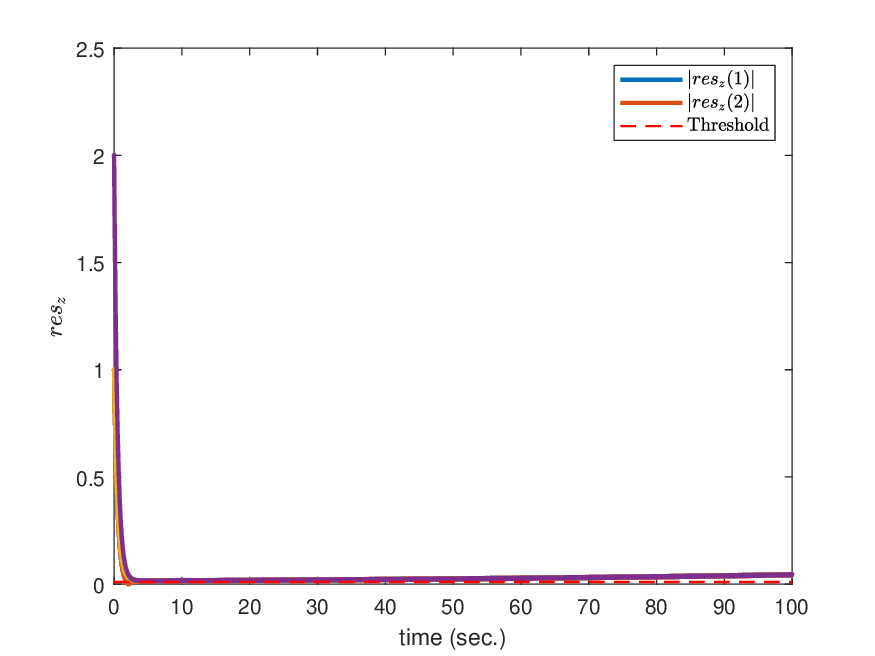}    
		\caption{The residual $res_z$ when the system is subject to ZD cyber-attack on the input channel and the cyber-attack signal $a_z$ is designed based on (\ref{eq:CovertAz}). As demonstrated, the residual will not exceed the threshold.}
		\label{fig:Residual1}
	\end{center}
\end{figure}

\begin{figure}
	\begin{center}
		\includegraphics[width=8.6cm]{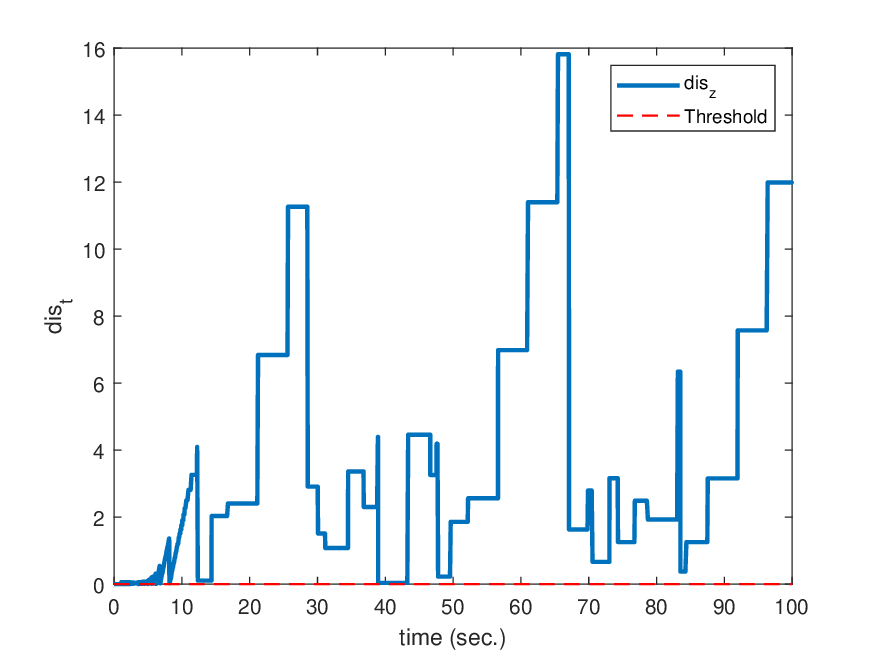}    
		\caption{The residual (\ref{eq:Res_t}) when the system is under the ZD cyber-attack where the residual exceeds the threshold and the cyber-attack is detected. }
		\label{fig:Residual}
	\end{center}
\end{figure}

We consider three different scenarios. In our first scenario, we consider that the attacker injects the ZD cyber-attack signal $a_u$ without launching any cyber-attacks on the auxiliary channel. This case is similar to the scenario in \cite{baniamerian2020monitoring} where no cyber-attacks occur on the auxiliary channel. The residual $res_z$ can be seen in Fig. \ref{fig:Residual1} where the ZD cyber-attack caused the residual to slowly increase and exceed the threshold. Furthermore, $dis_t$ will also exceed the threshold as demonstrated in Fig. \ref{fig:Residual}.

\begin{figure}
	\begin{center}
		\includegraphics[width=8.6cm]{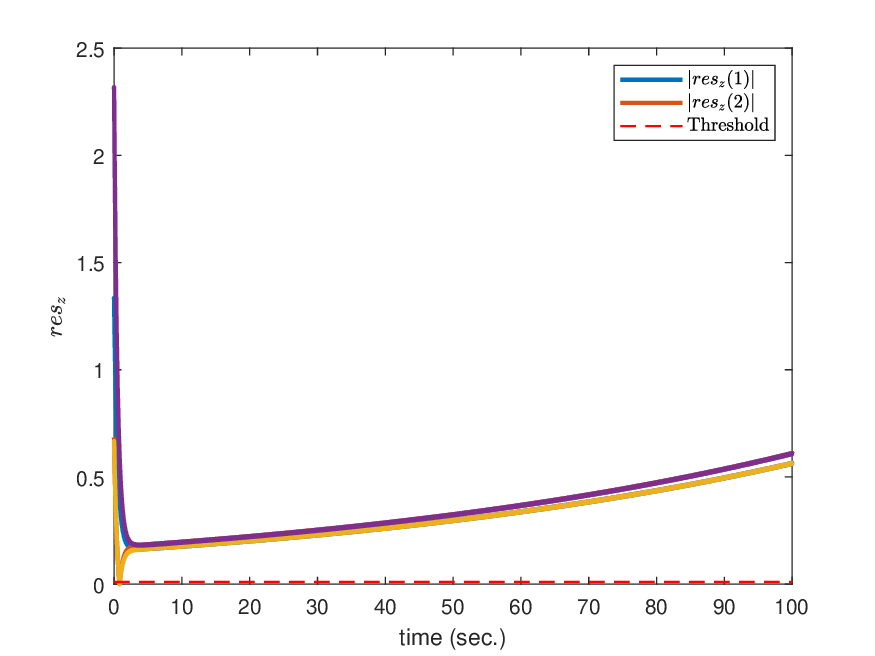}    
		\caption{The residual $res_z$ when the system is subject to ZD cyber-attack on the input channel and the cyber-attack signal $a_z$ is designed based on (\ref{eq:CovertAz}). As demonstrated, the residual will not exceed the threshold.}
		\label{fig:Residual3}
	\end{center}
\end{figure}

In the second scenario, we consider that the attacker launches FDI cyber-attacks on the auxiliary channel. In this case, we consider that the attacker is not aware of the dynamics of the auxiliary system. Since the attacker is unaware of the auxiliary system dynamics, the cyber-attack on the auxiliary channel is considered to be $a_z(t)=-a_u(t)$. Figure \ref{fig:Residual3} demonstrates the residuals in this scenario and as shown, the residual $res_z$ will exceed the threshold. Note that in this scenario, $dis_t$ will be the same as the first scenario. This is due to the fact that cyber-attacks on the auxiliary channel will not affect the triggering time of the auxiliary channel.

\begin{figure}
	\begin{center}
		\includegraphics[width=8.6cm]{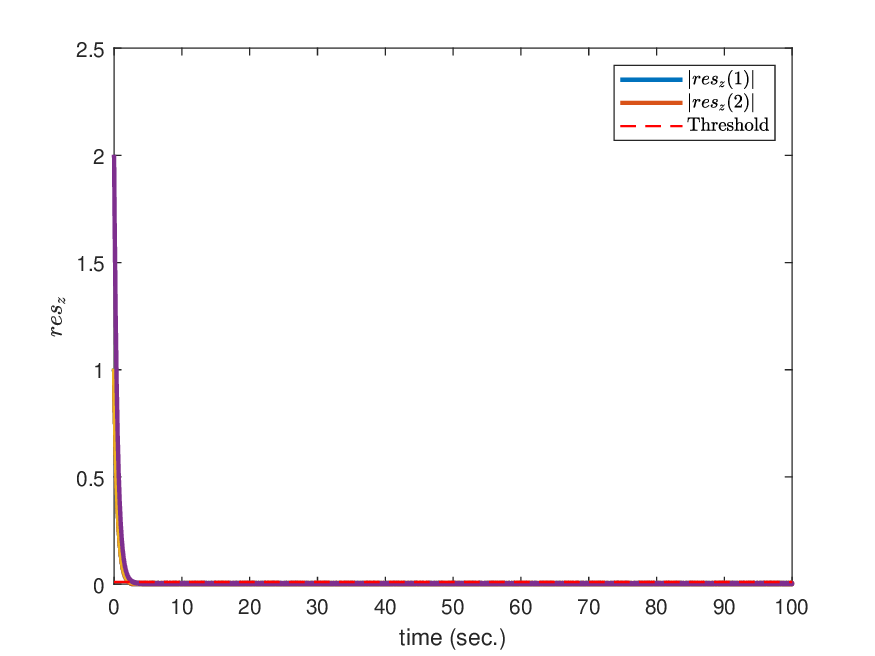}    
		\caption{The residual $res_z$ when the system is subject to ZD cyber-attack on the input channel and the cyber-attack signal $a_z$ is designed based on (\ref{eq:CovertAz}). As demonstrated, the residual will not exceed the threshold.}
		\label{fig:Residual2}
	\end{center}
\end{figure}

Finally, in our third scenario, the attacker injects $a_z$ by using (\ref{eq:CovertAz}). Therefore, the residual $res_z$ in (\ref{eq:Res_z}) will not exceed the threshold as shown in Fig. \ref{fig:Residual2}. However, $dis_t$ in (\ref{eq:Res_t})  will exceed its threshold similar to previous scenarios as depicted in Fig. \ref{fig:Residual} when the system is subject to the ZD cyber-attack, indicating that our detection mechanism is capable of detecting the ZD cyber-attack even when the attacker knows the dynamics of the auxiliary system and can launch a coordinated FDI attack (i.e., covert attack) on the auxiliary channel by injecting $a_z$.

\begin{figure}
	\begin{center}
		\includegraphics[width=8.6cm]{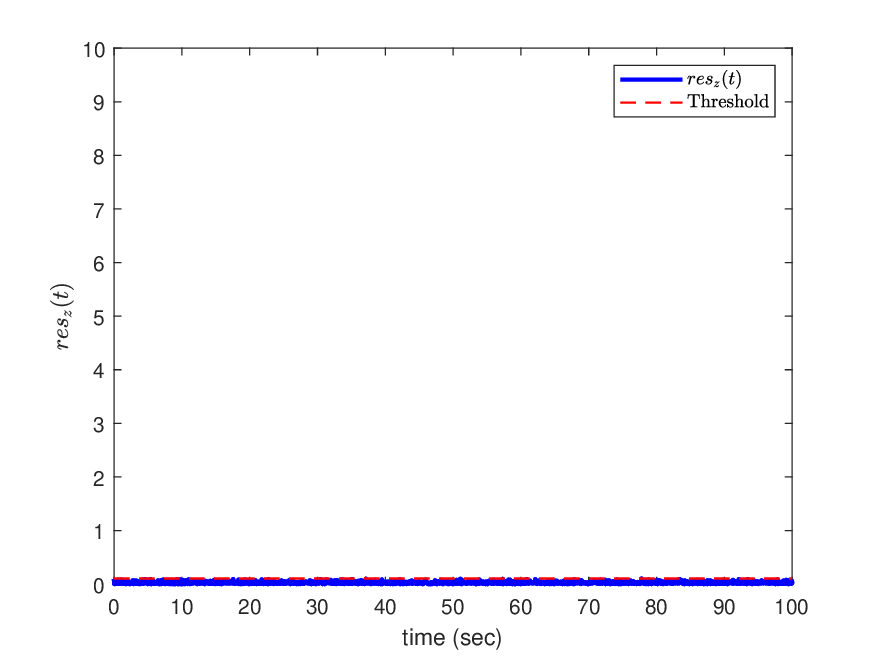}    
		\caption{The residual $res_z$ in Case II when the aircraft is subject to ZD cyber-attack on the input channel and the cyber-attack signal $a_z$ is designed based on (\ref{eq:CovertAz}). As demonstrated, the residual will not exceed the threshold.}
		\label{fig:NormalScenarioDis_t}
	\end{center}
\end{figure}

\begin{figure}
	\begin{center}
		\includegraphics[width=8.6cm]{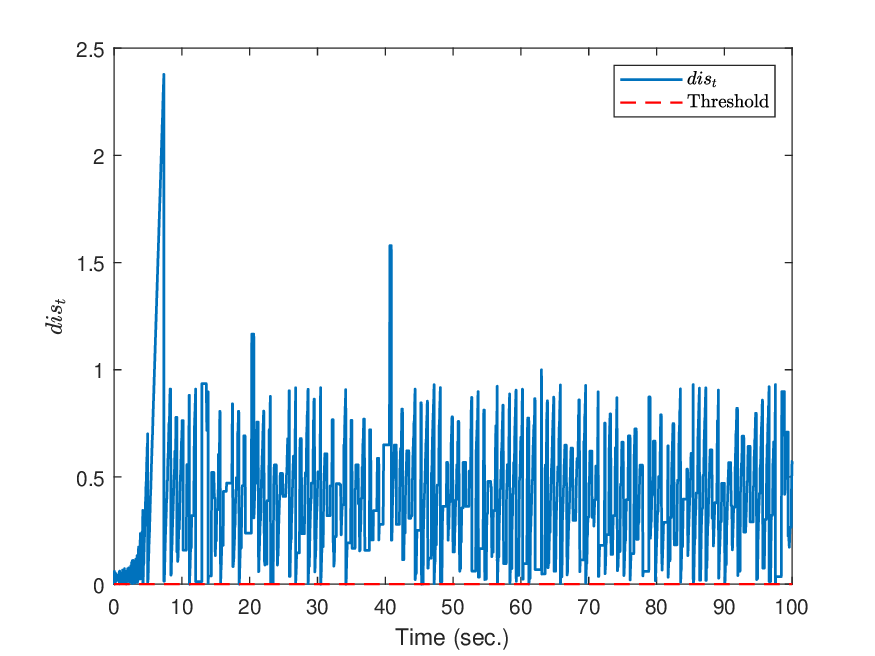}    
		\caption{The discrepancy $dis_t$ in Case II when the aircraft is subject to ZD cyber-attack. As demonstrated, the residual exceeds the threshold and the ZD cyber-attack is detected.}
		\label{fig:AircraftResidual}
	\end{center}
\end{figure}

\textbf{Case II}: In this case, we consider the linearized throttle to speed dynamics of an aircraft when the elevator is fixed as provided in \cite{stevens2015aircraft}. Note that in this system, there exists a non minimum phase zero at $s_0=0.0601$, and with $x_0=\begin{bmatrix}
	0.4327\\ 0.9001
\end{bmatrix}$, one can obtain $a_0=0.0507$. We consider the constant gains for the event triggering conditions similar to Case I with $K=[0.7499,    3.9509]$ and $L=9$. Figure \ref{fig:NormalScenarioDis_t} demonstrates the residual $res_z$ when the system is subject to ZD cyber-attack and the auxiliary channel is subject to $a_z$ as designed in (\ref{eq:CovertAz}). As demonstrated, the residual in this scenario will not exceed the threshold. Furthermore, Fig. \ref{fig:AircraftResidual} demonstrates the discrepancy $dis_t$ when the system is subject to ZD cyber-attack. As demonstrated, the discrepancy will exceed the threshold and the ZD cyber-attack is detected.

\section{Conclusion}
\label{sec:Conclusion}
In this paper, we have developed a ZD cyber-attack detection mechanism by using auxiliary systems with self-triggering communication channels. We have constructed residuals capable of detecting the ZD cyber-attacks even when the attacker know the auxiliary system dynamics and can launch cyber-attacks on all the communication channels.
In our future studies, we aim to extend our results to the ZD cyber-attacks in delayed and nonlinear CPS. Finally, considering a multi-agent framework is another promising future direction.

\bibliographystyle{IEEEtran}
\bibliography{sample}  
\end{document}